\documentclass[11pt,a4wide]{article}
\pdfoutput=1
\usepackage{jheppub}
 \usepackage{graphicx}
 \usepackage{bm}
 \usepackage{braket}
\usepackage{epsf,
shadow,pifont}
 \usepackage{amsmath,epsfig}
 \usepackage{amssymb,amsfonts,amsthm,amstext,amscd,dsfont}
\usepackage{latexsym}
\usepackage{slashed}
\usepackage{float}
\usepackage{epsfig}
\usepackage{simplewick}

\def\be{\begin{equation}}
\def\ee{\end{equation}}
\def\ba{\begin{eqnarray}}
\def\ea{\end{eqnarray}}
\def\D{\Delta}
\def\t{\tau}
\def\a{\alpha}
\def\b{\beta}

\def\D{\Delta}

\def\g{\gamma}
\def\G{\Gamma}

\def\nn{\nonumber\\}
\def\pa{\partial}
\def\s{\sigma}
\def\e{\epsilon}

\def\vp{\varphi}

\title{Lightcone expansions of conformal blocks in closed form}

\author{Wenliang Li}
\affiliation{Okinawa Institute of Science and Technology Graduate University, 1919-1 Tancha, Onna-son, Okinawa 904-0495, Japan}

\emailAdd{lii.wenliang@gmail.com}

\abstract{
We present new closed-form expressions for 4-point scalar conformal blocks in the s- and t-channel lightcone expansions. 
Our formulae apply to intermediate operators of arbitrary spin in general dimensions. 
For physical spin $\ell$, they are composed of at most $(\ell+1)$ Gaussian hypergeometric functions at each order. 
}
\preprint{}

\begin{document}

\maketitle 

\section{Introduction}
Conformal blocks are fundamental to the conformal bootstrap program \cite{Ferrara:1973yt, Polyakov:1974gs}, 
which aims to classify and solve conformal field theories (CFTs) by general consistency requirements, such as associativity of operator product expansion (OPE). 
In two dimensions, the conformal symmetry is infinite-dimensional, 
and the conformal bootstrap program can be carried out rather successfully \cite{Belavin:1984vu}. 
The well-known examples include the minimal models, which describe the critical phenomena of 2d statistical models, 
such as the Lee-Yang,  Ising, and Potts models \cite{DiFrancesco:1997nk}. 
On the other hand, the $d>2$ conformal bootstrap is considerably more challenging 
as the conformal symmetry is usually finite-dimensional. 

When performing operator product expansions in conformal field theories, 
the contributions of a primary and its descendants are related by conformal symmetry. 
A conformal block is defined by the contributions of a full conformal multiplet, 
which includes a primary and infinitely many descendants. 
The study of conformal blocks has a long history 
\cite{Ferrara:1973vz, Ferrara:1974nf, Ferrara:1974ny, Dobrev:1975ru, Lang:1992zw}, 
which dates back to the 1970's when the conformal bootstrap proposal was just formulated. 
The understanding of conformal blocks was significantly advanced by the works of Dolan and Osborn \cite{Dolan:2000ut, Dolan:2003hv, Dolan:2011dv}, 
in which they found explicit analytic expressions in $d=2,4,6$ dimensions, 
recursion relations and Casimir differential equations in general dimensions. 
These results had paved the road for the revival of the $d>2$ conformal bootstrap \cite{Rattazzi:2008pe}, 
where a new numerical conformal bootstrap method was proposed. 
Let us mention here the precise determinations of 3d Ising critical exponents \cite{ElShowk:2012ht, El-Showk:2014dwa, Kos:2014bka, Kos:2016ysd}, 
but refer to \cite{Poland:2018epd} for a comprehensive review on many other impressive results. 
\footnote{See also \cite{Qualls:2015qjb, Rychkov:2016iqz, Simmons-Duffin:2016gjk, Chester:2019wfx} for useful lecture notes. }

For the numerical conformal bootstrap, it may be adequate to be able to evaluate conformal blocks efficiently 
using the Zamolodchikov-like recursion relations \cite{Zamolodchikov:1987, Kos:2013tga, Penedones:2015aga} 
in the radial coordinates \cite{Pappadopulo:2012jk,Hogervorst:2013sma}. 
For the analytic conformal bootstrap, it is more desirable to have closed-form expressions of conformal blocks to enable general analytic computations, 
with the dream in mind that the 3d Ising CFT will eventually be solved analytically. 
Besides those mentioned earlier, 4-point conformal blocks in position space have been studied in various approaches 
\cite{Costa:2011dw,SimmonsDuffin:2012uy, Fitzpatrick:2013sya,  Hogervorst:2013kva, Costa:2014rya, Echeverri:2015rwa, Iliesiu:2015qra, Rejon-Barrera:2015bpa, Iliesiu:2015akf, Echeverri:2016dun,Isachenkov:2016gim, Fortin:2016lmf, Costa:2016hju, Costa:2016xah, Hogervorst:2016hal, Chen:2016bxc,  Karateev:2017jgd, Schomerus:2017eny, Isachenkov:2017qgn, Kravchuk:2017dzd, Fortin:2019fvx, Li:2019dix, Erramilli:2019njx, Buric:2019dfk}. 
In this paper, we will focus on conformal blocks with external scalars. In principle, spinning conformal blocks can be obtained from the scalar case using differential operators. 

In the analytic conformal bootstrap, considerable recent efforts have been devoted to solving the crossing equations near the lightcone, 
where the correlator is dominated by the contributions of low twist operators. 
\footnote{Other active analytic approaches include the Polyakov-Mellin bootstrap \cite{Gopakumar:2016wkt,Gopakumar:2016cpb,Dey:2016mcs,Dey:2017oim,Gopakumar:2018xqi}, analytic functionals \cite{Mazac:2016qev,Mazac:2018mdx,Mazac:2018ycv,Paulos:2019gtx,Mazac:2019shk}, and Tauberian theorems \cite{Qiao:2017xif,Mukhametzhanov:2018zja}. }
The crossing equations follow from OPE associativity when applied to 4-point functions of scalar primaries. 
By considering the double lightcone limit, one can show that the leading contribution from the identity operator indicates 
the existence of infinitely many higher spin operators whose twists are asymptotic to the sum of two external scaling dimensions 
\cite{Fitzpatrick:2012yx, Komargodski:2012ek}.
\footnote{We assume that the vacuum state has the lowest twist and the twist spectrum has a gap. 
In 2d CFTs the twist-0 spectrum is usually degenerate due to the vacuum Virasoro module. 
See \cite{Kusuki:2018wpa, Collier:2018exn} for the lightcone bootstrap in 2d. }
The double-twist phenomena were noticed earlier in a more concrete context \cite{Alday:2007mf}.  
The same argument can be extended to subleading twists, but one needs to be more careful about the potential mixing of different twist families. 
In some sense, these large spin sectors behave like mean fields or generalized free fields. 
Then the next step is to study the systematic corrections to the ``free" theories 
\cite{Vos:2014pqa,Kaviraj:2015cxa,Alday:2015eya,Kaviraj:2015xsa,Alday:2015ota,Alday:2015ewa,Alday:2016mxe,Alday:2016njk,Alday:2016jfr,Simmons-Duffin:2016wlq,Dey:2017fab,Sleight:2018epi,Cardona:2018dov}. 
Naturally, one can formulate the problem as a perturbation theory in large spin. 
As in many perturbation problems,  
the results are asymptotic series in the expansion parameter, i.e. $1/\ell$ \cite{Alday:2015ewa}. 
However, it turned out that the results for the leading twist family of the 3d Ising CFT are consistent with the numerical values down to spin two \cite{Alday:2015ota,Simmons-Duffin:2016wlq}. 
The lightcone bootstrap should admit a convergent formulation. 

The asymptotic issue was later resolved by the elegant Lorentzian inversion formula proposed in \cite{Caron-Huot:2017vep}, 
which upgrades the asymptotic expansion to a convergent integral transform 
and establishes the analyticity in spin assumed earlier \cite{Costa:2012cb}. 
\footnote{In \cite{Sleight:2018ryu}, convergent results were obtained from a different approach.}       
An alternative derivation was presented in \cite{Simmons-Duffin:2017nub}  
and a generalization to the spinning case was given in \cite{Kravchuk:2018htv}. 
In two and four dimensions, general closed-form results of the Lorentzian inversion of a conformal block can be found in \cite{Liu:2018jhs}. 
In $d\neq 2,4$ dimensions, the nonperturbative Lorentzian inversion in the lightcone limit has also been studied in \cite{Cardona:2018qrt,Albayrak:2019gnz}\cite{Li:2019dix}, 
which encodes the OPE data of the leading twist spectrum. 
Interestingly, it was shown that the analyticity in spin extends to spin-0 in some cases \cite{Turiaci:2018dht, Alday:2017zzv,Henriksson:2018myn,Alday:2019clp}, 
despite the presence of a singularity between $\ell=0$ and $\ell=2$ related to a poor Regge limit.  
More recently, a closely related dispersion relation for conformal field theory was proposed in \cite{Carmi:2019cub}, 
which expresses a correlator as an integral over the double discontinuity. 
\footnote{See also \cite{Bissi:2019kkx} for another recently proposed dispersion relation, which is based on single discontinuity. }

In this paper, we will focus on the 4-point functions of scalar primaries: 
\be
\langle \vp_1\,\vp_2\,\vp_3\,\vp_4\rangle
=
\bigg(\frac {x_{24}}{x_{14}}\bigg)^{2a}\,
\bigg(\frac {x_{14}}{x_{13}}\bigg)^{2b}\,
\frac {\mathcal G(u,v)} {x_{12}^{\D_1+\D_2}\, x_{34}^{\D_3+\D_4}}\,,
\label{4-p-fn}
\ee
where $x_i$ indicates the positions of the external scalars $\vp_i$, 
and $x_{ij}=|x_i-x_j|$ is the distance between two operators. 
In general, the external operators can have different scaling dimensions, 
and their differences are denoted by
\be
a=\frac {\D_1-\D_2} 2\,,\quad
b=\frac{\D_3-\D_4} 2\,.
\ee
The functional form of the 4-point function is determined by conformal symmetry up to a function of 
two conformally invariant cross-ratios
\be
u=\frac {x_{12}^2\, x_{34}^2}  {x_{13}^2\, x_{24}^2},\quad
v=\frac {x_{14}^2\, x_{23}^2}  {x_{13}^2\, x_{24}^2}\,.
\ee 
In the Lorentzian signature, 
the first cross-ratio $u$ vanishes when the pair of operators in the s-channel OPE, i.e. $\vp_1\times\vp_2$, are light-like separated, 
so we will call $u\rightarrow 0$ the s-channel lightcone limit. 
Analogously, $v\rightarrow 0$ will be called the t-channel lightcone limit, which is related to $\vp_2\times\vp_3$. 
Usually, correlators develop singularities in the lightcone limits, 
which are associated with ambiguities in time ordering when operators are time-like separated.
It is sometimes convenient to switch to $(z,\bar z)$, which are real, independent coordinates in the Lorentzian signature. 
\footnote{In the Euclidean signature, $z,\bar z$ are complex conjugate to each other. 
In terms of $u,v$, the boundary between the Lorentzian and Euclidean regimes is given by $(1+u-v)^2=4\,u$.  }
They are related to $(u, v)$ by 
\be
u=z\bar z\,,\quad v=(1-z)(1-\bar z)\,.
\label{zzb-definition}
\ee
It is straightforward to obtain the lightcone expansions in terms of $z,\bar z$ from those in terms of $u,v$ via \eqref{zzb-definition}.

For an intermediate scalar, i.e. $\ell=0$, the double lightcone expansion of a 4-point scalar conformal block is known in closed form in general dimensions
\footnote{This is the main reason that we use $u,v$ rather than $z,\bar z$.}, 
given by hypergeometric functions of two variables \cite{Dolan:2000ut}:
\ba
G^{(d,a,b)}_{\D,0}(u,v)&=&
u^{\D/2}\,v^{\frac{a-b}{2}}\bigg[
\frac{\G(b-a)}{(\D/2)_{-a}\,(\D/2)_b}\,v^{\frac{a-b}2}
\sum_{k_1,\,k_2=0}^\infty
C^{(\ell=0)}_{k_1,\,k_2}\,u^{k_1}v^{k_2}
+(a\leftrightarrow b)
\bigg]\,,
\label{lightcone-expansion-scalar}
\ea
where the series coefficients take the form
\be
C^{(\ell=0)}_{k_1,k_2}=\frac{1}{k_1!\,k_2!}\,\frac{(\D/2+a)_{k_1+k_2}\,(\D/2-b)_{k_1+k_2}}{(\D-d/2+1)_{k_1}\,(1+a-b)_{k_2}}\,.
\ee
The two-variable hypergeometric functions belong to Appell's function of type $F_4$. 
One can carry out the $k_1$ or $k_2$ summation, 
which corresponds to a hypergeometric function of type ${}_2F_1$. 
Therefore, at each order of the s- or t-channel lightcone expansion, the dependence on the other cross-ratio is encoded in one Gaussian hypergeometric function. 

The main goal of this paper is to extend \eqref{lightcone-expansion-scalar} to the case of generic $\ell$. 
For physical spin $\ell$, 
we find that the s- and t-channel lightcone expansions of a generic 4-point scalar conformal block read: 
\ba
G^{(d,a,b)}_{\t,\ell}(u,v)&\sim& u^{\t/2}\,v^{\frac{a-b} 2}\bigg[v^{\frac{a-b} 2}
\sum_{k=0}^\infty u^{k}
\sum_{n=0}^{\ell}  A_{k,n}\,(1-v)^{\ell-n}
{}_2F_1[\dots, v]+(a\leftrightarrow b)\bigg],
\label{small-u-schematic}
\\
G^{(d,a,b)}_{\t,\ell}(u,v)
&\sim& u^{\t/2}\,v^{\frac{a-b} 2}\bigg[v^{\frac{a-b} 2}
\sum_{k=0}^\infty v^{k}
\sum_{n=0}^\ell B_{k,n}\,
(1-u)^{\ell-n}\,u^{n}
{}_2F_1[\dots, u]+(a\leftrightarrow b)\bigg],
\label{small-v-schematic}
\ea
where for simplicity some normalization factors and the parameters in the ${}_2F_1$ hypergeometric functions have been omitted. 
As generalizations of the $\ell=0$ case, 
the dependence on the other cross-ratio is now encoded in at most $(\ell+1)$ Gaussian hypergeometric functions. 
The precise formulae and closed-form coefficients can be found in 
(\ref{u-expansion-1-v}-\ref{f0}, \ref{Big-A})
and (\ref{gkn}, \ref{small-v-solution-explicit}, \ref{big-B}, \ref{gkn-2}). 
The same formulae also apply to the case of generic $\ell$,
but the $n$-summation should be from $0$ to $2k$ for \eqref{small-u-schematic} and from $0$ to $\infty$ for \eqref{small-v-schematic}. 
The double lightcone expansion can be obtained from either \eqref{small-u-schematic} or \eqref{small-v-schematic}: 
\be
G^{(d,a,b)}_{\t,\ell}(u,v)\sim u^{\t/2}\,v^{\frac{a-b} 2}\bigg[v^{\frac{a-b} 2}
\sum_{k_1,\,k_2=0}^\infty
C_{k_1,\,k_2}\,u^{k_1}\,v^{k_2}
+(a\leftrightarrow b)\bigg]\,, 
\label{double-lightcone}
\ee
where the series coefficients $C_{k_1,k_2}$ are finite sums of $A_{k,n}$ in \eqref{BigC-2} or $B_{k,n}$ in \eqref{BigC-1}. 
The double power series \eqref{double-lightcone} furnishes a general, explicit solution to 
the quadratic and quartic Casimir equations, which are partial differential equations of two variables.  

For comparison, we will briefly discuss below other complete analytic formulae of 4-point scalar conformal blocks in general dimensions
: 
\begin{itemize}
\item
According to \eqref{zzb-definition}, the conformal blocks are symmetric functions in $z,\bar z$, 
so it is natural to expand the conformal blocks in terms of building blocks that are manifestly symmetric in $z,\bar z$. 
In \cite{Dolan:2003hv}, the conformal blocks with non-negative integer spin $\ell$ are given by an infinite sum of 
Jack polynomials, which are symmetric polynomials in $z, \bar z$. 
The coefficients are presented in closed-form in terms of ${}_4F_3$ hypergeometric series. 
More explicitly, the two-variable Jack polynomials can be expressed as the Gegenbauer polynomials in $(z+\bar z)/(2\sqrt{z \bar z}\,)$ times a power function of $z \bar z$. However, this formula is not very convenient for the lightcone expansions considered in this paper. 
For instance, the lightcone limits receive contributions from 
infinitely many Jack polynomials.

\item
In Mellin space, 4-point scalar conformal blocks in general dimensions are also known in closed form for non-negative integer $\ell$. 
The essential part is known as the Mack polynomials \cite{Mack:2009mi} \cite{Dolan:2011dv}, which are the counterparts of $A_{k,n}, B_{k,n}$ in position space 
\footnote{The Mack polynomials are related to the continuous Hahn polynomials \cite{Gopakumar:2016wkt,Gopakumar:2016cpb}. Similarly, $A_{k,n}, B_{k,n}$ may be related to some orthogonal polynomials. }. 
Due to our choices of basis functions, 
$A_{k,n}, B_{k,n}$ are much simpler than the results from the Mack polynomials at low orders of the lightcone expansions. 
In addition, $A_{k,n}, B_{k,n}$ directly apply to generic $\ell$, as opposed to only non-negative integer $\ell$ in the case of the Mack polynomials. 
\footnote{
See \cite{Chen:2019gka} for a recent generalization to continuous spin.
} 
\item
As shown in \cite{Isachenkov:2016gim}, 
the quadratic Casimir equation can be mapped to the Schr\"odinger equations for integrable Calogero-Sutherland models. 
Then 4-point scalar conformal blocks are related to the Harish-Chandra functions, 
which can be written as double infinite sums of Gaussian hypergeometric functions in $z,\bar z$ with power law insertions 
\cite{Isachenkov:2017qgn}. 
To the best of our knowledge, 
while it is straightforward to derive the s-channel lightcone expansion, 
i.e. $z\ll1$, from the explicit results in \cite{Isachenkov:2017qgn}, 
it is more difficult to obtain compact formulae for the t-channel lightcone expansion, i.e. $1-\bar z\ll 1$. 
Our s-channel lightcone expansion also seems to be simpler, at least at low orders. 
As solutions of the quadratic Casimir equation, the formulae in this paper provide alternative expressions for the Harish-Chandra functions 
in the small $z \bar z$ or $(1-z)(1-\bar z)$ expansion. 
\end{itemize}

\section{Lightcone limits of conformal blocks}
In this section, we will give a brief overview of 4-point scalar conformal blocks in the lightcone limits. 
After performing the s-channel operator product expansion, 
the conformal invariant part decomposes into s-channel conformal blocks 
\be
\mathcal G(u,v)=\sum_i\, P_i \, G^{(d,a,b)}_{\t_i,\ell_i}(u,v)\,,
\ee
where $P_i$ is the product of two OPE coefficients associated with the intermediate primary operator $\mathcal O_i$, 
and $(\t_i, \ell_i)$ are the twist and spin of $\mathcal O_i$. 
Note that the primary operator is labeled by twist
\be
\t=\D-\ell\,,
\ee
rather than the scaling dimension $\D$, 
because twist appears naturally in the lightcone limit. 
The s-channel conformal block $G^{(d,a,b)}_ {\t_i,\ell_i}(u,v)$ encodes the contributions of $\mathcal O_i$ and its descendants in the s-channel OPE, 
and satisfies the quadratic Casimir differential equation
\be
\mathcal D\,\bar G=
\frac 1 2 \big[(\t+\ell)(\t+\ell-d)+\ell(\ell+d-2)\big]\,
\bar G\,,
\label{Casimir-eq}
\ee
where
\ba
\quad
\mathcal D=2D_u^2-dD_u+\frac{1-u-v}{v}\Bigg[D_v^2-\bigg(\frac {a-b} {2}\bigg)^2\Bigg]
-(1+u-v)\Bigg[(D_u+D_v)^2-\bigg(\frac {a+b} {2}\bigg)^2\Bigg]
\,,\nn
\label{Casimir-D}
\ea
and 
\be
\bar G=v^{-\frac{a-b} 2}\, G^{(d,a,b)}_{\t,\ell}(u,v),\quad
D_u=u\,\pa_u,\quad
D_v=v\,\pa_v\,.
\ee
As the differential operator $\mathcal D$ is symmetric in $a$ and $b$, 
two independent solutions can be related by interchanging $a$ and $b$. 
The boundary condition is given in the short distance limit: 
\be
G^{(d,a,b)}_{\t,\ell}(u,v)=\frac{\G(\t/2+\ell)^2}{\G(\t+2\ell)}\,u^{\t/2}\,(1-v)^\ell, \quad 
\text{as} \quad u\rightarrow 0, \,v\rightarrow 1\,.
\ee
For physical operators, the spin $\ell$ is a non-negative integer, i.e. $\ell=0,1,2,\dotsb$. 
We will consider conformal blocks with generic $\ell$ using the Casimir equation \eqref{Casimir-eq}, which can be considered as analytic continuation of the physical conformal blocks. 
Operators of generic spin $\ell$ are nonlocal. 
They play a significant role in the Lorentzian inversion formula \cite{Caron-Huot:2017vep}
and can be understood as a combination of the spin-shadow and light transforms of local operators \cite{Kravchuk:2018htv}. 
Analogous to the shadow transform, these transforms can be viewed as Weyl reflections. 
From the perspective of conformal Casimirs, the eigenvalues are invariant under the Weyl reflections
\footnote{Here the independent labels are $(\D,\ell)$. 
These Weyl reflections are not independent. } 
\be
\D\leftrightarrow d-\D\,,\quad
\ell\leftrightarrow 2-d-\ell\,,\quad
\D\leftrightarrow 1-\ell\,,
\ee
which correspond to the shadow, spin-shadow and light transforms. 
Note that the spin-shadow transform also appears in the closed-form expressions of conformal blocks in 
even dimensions. 

In the s-channel lightcone limit $u\rightarrow 0$, the coefficients of $D_u, D_v$ in \eqref{Casimir-D} are regular, 
so we can substitute $D_u$ with the leading exponent $\t/2$ and set $u$ to $0$. 
Then we obtain a second order differential equation in $v$ for the leading term, whose explicit solution reads:
\be
G^{(d,a,b)}_{\t,\ell}(u,v)\big|_{u\rightarrow 0}=\frac{\G(\t/2+\ell)^2}{\G(\t+2\,\ell)}\,u^{\t/2}\,(1-v)^\ell\,
{}_2F_1\bigg[\begin{matrix}
\t/2+\ell-a,\t/2+\ell+b\\
\t+2\,\ell
\end{matrix}\,;1-v\bigg]\,.
\label{small-u-limit}
\ee
The conformal blocks are normalized such that the double lightcone limit with $a=b=0$ is
\be
G^{(d,0,0)}_{\t,\ell}(u,v)\big|_{u,v\rightarrow 0}= -\,u^{\t/2}\,\big(\log v+2 H_{\t/2+\ell-1}\big)\,,
\ee 
where $H_x=H(x)$ is the Harmonic number. 

In the t-channel lightcone limit $v\rightarrow 0$, the quadratic Casimir equation does not reduce to an equation for the leading term, 
but one can make use of both the quadratic and quartic Casimir equations to derive a closed equation \cite{Caron-Huot:2017vep}. 
As the resulting equation is of fourth order, it is more challenging to obtain explicit solutions, 
in contrast to the standard second order equation from the small $u$ limit. 
Nevertheless, 
a general closed-form expression for the small $v$ limit was recently found in \cite{Li:2019dix}:
\ba
&&G^{(d,a,b)}_{\t,\ell}(u,v)\big|_{v\rightarrow 0}
=
v^{\frac{a-b} 2}\,
\Bigg[\frac{\G(b-a)\,}{(\t/2+\ell)_{-a}\,(\t/2+\ell)_{b}}\,\frac{u^{\t/2}\,v^{\frac{a-b} 2}}{ (1-u)^{(d-2)/2+(a-b)}}\,
\nn&&\qquad\qquad\qquad\qquad\qquad\times\,
F_{0,2,1}^{0,2,2}\bigg[\Big|
\begin{matrix}
-\ell, 3-d-\ell
\\
\g, 2-d/2-\ell
\end{matrix}\,\Big|\,
\begin{matrix}
\g/2-a, \g/2+b
\\
\g/2+\t/2+\ell
\end{matrix}\,\Big|\, u, -u
\bigg]
+(a\leftrightarrow b)
\Bigg]\,,\qquad\quad
\label{small-v-solution}
\ea
where
\be
\g=\t-d+2\,,\quad
(x)_y=\frac{\G(x+y)}{\G(x)}\,.
\ee
For $\ell>0$, the anomalous dimension is given by $\g$. 
It is straightforward to verify that \eqref{small-v-solution} solves the quartic differential equation. 
Here we have introduced a two-variable hypergeometric function, the Kamp\'e de F\'eriet function. 
In our notation, the function is defined as
\begin{eqnarray}
&&F^{0,2,2}_{0,2,1}
\bigg[\Big|
\begin{matrix}
\a_1, \a_2\\
\b_1,\b_2
\end{matrix}
\Big|
\begin{matrix}
\a_3, \a_4\\
\b_3
\end{matrix}
\Big|
x, y
\bigg]
=\sum_{m,n=0}^\infty\,
 \frac{(\a_1)_n\,(\a_2)_n\,}{(\b_1)_n\,(\b_2)_n}\,
  \frac{(\a_3)_{m+n}\,(\a_4)_{m+n}}{(\b_3)_{m+n}}\,
  \frac{x^m\,y^n}{m!\,n!}\,.
 \qquad
\end{eqnarray}
Note that our definition is not standard. Usually the terms with $(m+n)$ are on the left of those with $m$ or $n$. 
In our notation, it is more clear that the $n$-summation terminates for physical spin $\ell$ due to the Pochhammer symbol $(-\ell)_n$. 
Then for each $n$ the $m$-summation corresponds to a ${}_2F_1$ hypergeometric function. 
More explicitly, we can write \eqref{small-v-solution} as
\be
G^{(d,a,b)}_{\t,\ell}(u,v)\big|_{v\rightarrow 0}
=
v^{\frac{a-b} 2}\,\Bigg[\frac{\G(b-a)\,}{(\t/2+\ell)_{-a}\,(\t/2+\ell)_{b}}\,v^{\frac{a-b} 2}\,
\sum_{n=0}^\ell\,B_{0,n}\, g_{0,n}(u)\,+(a\leftrightarrow b)
\Bigg]\,,
\ee
where the factorized coefficients are
\ba
B_{0,n}=\frac{\ell!}{n!(\ell-n)!}\,\frac{(3-d-\ell)_n}{(\g)_n\,(2-d/2-\ell)_n}
\frac{(\g/2-a)_n\,(\g/2+b)_n}{(\g/2+\t/2+\ell)_n}\,,
\label{B0n}
\ea
and the basis functions are
\ba
\quad g_{0,n}(u)=\frac{u^{\t/2+n}}{(1-u)^{(d-2)/2+(a-b)}}\,
{}_2F_1\left[
\begin{matrix} 
\g/2+n-a,\g/2+n+b \\
\g/2+n+\t/2+\ell
\end{matrix}
;\,u\right]\,.
\ea
The subscript $0$ indicates the lowest order in the small $v$ expansion. 

\section{Lightcone expansions of conformal blocks}
The lightcone limits of 4-point scalar conformal blocks are the leading terms of the lightcone expansions. 
In this section, we will consider the subleading terms. 
We will first generalize the formula for the $v\rightarrow 0$ limit to all orders in the small $v$ expansion. 
Then we will discuss the small $u$ expansion. 
The double lightcone expansion can be readily derived from either of them. 

\subsection{The t-channel lightcone expansion}
\label{subsec-small-v}
In the $v\rightarrow 0$ limit, \eqref{lightcone-expansion-scalar} should be equivalent to \eqref{small-v-solution} with $\ell=0$. 
To show their equivalence, let us perform a linear transformation of the ${}_2F_1$ function associated with the $m$-summation
\ba
&&u^{\D/2}\,{}_2F_1\bigg[
\begin{matrix}
\D/2+k+a, \D/2+k-b
\\
\D/2+\g/2
\end{matrix};\, u
\bigg]\nn
&=&\frac {u^{\D/2}} {(1-u)^{(d-2)/2+(a-b)+2k}}\,
{}_2F_1\bigg[
\begin{matrix}
\g/2-k-a, \g/2-k+b
\\
\g/2+\D/2
\end{matrix};\, u
\bigg]\,,
\label{2F1-transformation}
\ea
which is precisely $g_{0,n}(u)$ with $\ell=n=0$ when $k=0$. The coefficient of ${}_2F_1$ also matches with $B_{0,n}$ in \eqref{B0n} with $\ell=n=0$. 
In \eqref{2F1-transformation}, we have intentionally preserved the $k$-dependence, 
which suggests the general basis functions should be
\ba
g_{k,n}(u)=\frac{u^{\t/2+n}}{(1-u)^{(d-2)/2+(a-b)+2k}}\,
{}_2F_1\left[
\begin{matrix} 
\g/2+n-k-a,\g/2+n-k+b \\ 
\g/2+n+\t/2+\ell
\end{matrix}
;\,u\right]\,,\quad
\label{gkn}
\ea
where $\g=\t-d+2$. 
For $\ell=0,1,2,\dots$, the small $v$ expansion of a generic conformal block then reads
\be
G^{(d,a,b)}_{\t,\ell}(u,v)
=
v^{\frac{a-b} 2}\bigg[\frac{\G(b-a)\,}{(\t/2+\ell)_{-a}\,(\t/2+\ell)_{b}}\,
\sum_{k=0}^\infty\,v^{\frac{a-b} 2+k}\,
\sum_{n=0}^\ell\,B_{k,n}\, g_{k,n}(u)\,+(a\leftrightarrow b)
\bigg]\,.
\label{small-v-solution-explicit}
\ee
It is not surprising that conformal blocks can be expanded in terms of $g_{k,n}(u)$, 
since $g_{k,n}(u)\sim u^{\t/2+n}+\dotsb$ where the leading exponent increases with $n$. 
The nice feature is that the $n$-summation always terminates at $n=\ell$ when $\ell$ is a non-negative integer, 
generalizing a property of the small $v$ limit to all the subleading terms. 
Note that for a generic $\ell$ the $n$-summation does not terminate for both the leading and subleading terms.   

The next step is to find a general expression for $B_{k,n}$. 
One can compute $B_{k,n}$ using the quadratic Casimir equation \eqref{Casimir-eq} order by order. 
For example, we find
\ba
B_{1,n}&=&
\frac{B_{0,n}}
{(\g/2+n-a-1)(\g/2+n+b-1)}
\nn&&\times
\bigg[(\g/2-a-1)(\g/2+b-1)\Big(-\ell+\frac{(\t/2+\ell+a)(\t/2+\ell-b)}{1+a-b}\Big)
\nn&&\qquad
-\frac n 2 \bigg(\frac{\t+\ell-1}{2-d-\ell+n}(d-2)(-a+b-d/2-1)
\nn&&\qquad\qquad\quad
-\frac{(\t+\ell-1)_2}{(1-d-\ell+n)_2}
\Big(2\ell(n-\ell)+(d-2)(1-2\ell+n)\Big)\bigg)
\bigg]\,,\quad
\ea
where $B_{0,n}$ is defined in \eqref{B0n}. 
In general, they are rational functions of $(\ell,\t,a,b,d)$. 
One can decompose them into factorized rational functions according to their poles, i.e. the zeros of the denominators. 
It turns out that $B_{k,n}$ can be expressed as
\ba
B_{k,n}
&=&\sum_{n_k=0}^k\,\sum_{n_i=0}^n\,
\frac{(-\ell)_{n_k+n_3}\,(\ell-n+1)_{n-n_1}\,(\frac{d-2} 2+n_3)_{n_1-n_3}\,(\t+\ell-1)_{n_2+n_3}\,}
{(k-n_k)!\,(n_k-n_2)!\,(n-n_1)!\,(n_1-n_2)!\,(n_2-n_3)!\,(n_2+n_3-n_1)!}
\nn&&\qquad\quad\times
\frac{(\t/2+\ell+a)_{k-n_k}\,(\t/2+\ell-b)_{k-n_k}}{(1+a-b)_{k-n_k}}\,
\Big(-a+b-\frac {d-2} 2-2k\Big)_{n_2-n_3}\,
\nn&&\qquad\quad\times
\frac{(3-d-\ell)_{n-n_1}}{(\g)_n\,(2-d/2-\ell)_n}\,
\frac{(\g/2+n_2-k-a)_{n-n_2}\,(\g/2+n_2-k+b)_{n-n_2}}{(\g/2+\t/2+\ell)_n}
\,,
\quad
\label{big-B}
\ea
where $n_i$ indicates $(n_1,\, n_2,\,n_3)$. 
To arrive at \eqref{big-B}, the poles from $1/(1+a-b)_{m}$ are particularly helpful. 
The last line of \eqref{big-B} also shares some nice features of the leading term in \eqref{B0n}, 
where several terms coincide with the parameters in the associated ${}_2F_1$ functions. 
Since $(p!)^{-1}$ vanishes if $p$ is a negative integer, there are additional conditions for non-vanishing summands
\be
n_k\geq n_2\,,\quad n_1\geq n_2\geq n_3,\quad n_2+n_3\geq n_1. 
\ee
The dependence on $n_k$ is relatively simple, so it is straightforward to carry out the $n_k$-summation. 
The $(n_1,n_2,n_3)$ are more entangled with each other, but $n_1$ can be eliminated easily as well. 
So there remain at most two summations. 
Although \eqref{big-B} is guessed from low order expressions, 
we have tested it at much higher orders using the Casimir equation \eqref{Casimir-eq}, 
which can be performed efficiently by setting $(\ell,\t,a,b,d)$ to rational numbers. 

Then we can derive the double lightcone expansion from \eqref{small-v-solution-explicit}:
\be
G^{(d,a,b)}_{\t,\ell}(u,v)
=
v^{\frac{a-b} 2}\,\Bigg[
\sum_{k_1,\,k_2=0}^\infty\,C_{k_1,\,k_2}\,u^{\t/2+k_1}\,v^{\frac{a-b} 2+k_2}+(a\leftrightarrow b)
\Bigg]\,,
\label{double-lightcone-expansion}
\ee
where the coefficients $C_{k_1,k_2}$ are sums of $B_{k,n}$ in \eqref{big-B}:
\ba
C_{k_1,k_2}&=&\frac{\G(b-a)\,}{(\t/2+\ell)_{-a}\,(\t/2+\ell)_{b}}\sum_{n,m=0}^{k_1}B_{k_2,n}\,\frac{(a-b+\frac{d-2}2+2k_2)_{k_1-n-m}}{(k_1-n-m)!}
\nn&&\qquad\qquad\qquad\qquad\qquad\qquad\times\,
\frac{( \g/ 2+n-k_2-a)_{m}\,( \g/ 2+n-k_2+b)_{m}}{m!\, (\g/2+n+\t/2+\ell)_{m}}\,.\quad
\label{BigC-1}
\ea

Before moving to the small $u$ expansion, let us discuss some properties of the building block $g_{k,n}(u)$. 
After a linear transformation, $g_{k,n}(u)$ becomes
\be
g_{k,n}(u)=u^{\t/2+n}\,(1-u)^{\ell-n}\,
{}_2F_1\left[
\begin{matrix} 
\t/2+\ell+k+a,\t/2+\ell+k-b \\ \t/2+\ell+\g/2+n
\end{matrix}
;\,u\right]\,.
\label{gkn-2}
\ee
The alternative expression \eqref{gkn-2} may look simpler than \eqref{gkn}, but the parameters of the ${}_2F_1$ function do not match with those in $B_{k,n}$. 
One may wonder whether the termination of the $n$-summation for non-negative integer $\ell$ is associated with the exponent of $(1-u)$. 
We will see that this is indeed the case.  

The boundary condition of conformal blocks is given in the limit $u\rightarrow 0, v\rightarrow 1$. 
The s-channel OPE in the $u\rightarrow 0$ limit is dominated by the contributions of low twist operators. 
Now we can also consider the dual limit $u\rightarrow 1$ of conformal blocks, 
which reduces to expanding $g_{k,n}(u)$ around $u=1$: 
\ba
g_{k,n}(u)&=&
\frac{\G(-1+a-b+d/2+\ell-n+2k)\,\G(\t+\ell-d/2+1+n)}
{\G(\t/2+\ell+k+a)\,\G(\t/2+\ell+k-b)}
\nn&&\quad\times\,
\frac{u^{\t/2+n}}{(1-u)^{(d-2)/2+(a-b)+2k}}\,{}_2F_1\bigg[
\begin{matrix}
\g/2+n-k-a,\g/2+n-k+b\\
2-a+b-d/2-\ell+n-2k
\end{matrix};\,1-u
\bigg]
\nn
&&+\,
\frac{\G(1-a+b-d/2-\ell+n-2k)\,\G(\t+\ell-d/2+1+n)}
{\G(\g+n-k-a)\,\G(\g+n-k+b)}
\nn&&\quad\times\,
u^{\t/2+n}\,(1-u)^{\ell-n}\,
{}_2F_1\bigg[
\begin{matrix}
\t/2+\ell+k+a,\t/2+\ell+k-b\\
a-b+d/2+\ell-n+2k
\end{matrix};\,1-u
\bigg]\,.\qquad
\ea 
The leading exponents are universal, which means that 
we are not able to organize the spectrum according to the asymptotic behaviour in the $u\rightarrow 1$ limit. 
In the first part, the leading exponent of $(1-u)$ is independent of $n$, 
which is in accordance with the facts that $g_{k,n}$ are natural basis functions for the lightcone expansion 
and $g_{0,n}$ was first found by stripping off this part in the small $u$ expansion. 
From the second part, we can see that the $n$-summation terminates naturally for physical spin $\ell$ when the leading exponent of $(1-u)$ becomes zero.

\subsection{The s-channel lightcone expansion}
\label{subsec-small-u}
Above, we discuss the small $v$ expansion. 
Now let us consider the small $u$ expansion. 
In general, the $v$-dependence can be encoded in 
$(2k+1)$ hypergeometric functions of type ${}_2F_1$ at order $u^{\t/2+k}$. 
As a generalization of the small $u$ limit in \eqref{small-u-limit}, 
we can expand a generic conformal block as 
\footnote{Similar expansions in terms of $z, \bar z$ were discussed in \cite{Simmons-Duffin:2016wlq}. 
Note that $A_{1,1}$ in \eqref{A-low-order} has a simpler expression than the counterpart $A_{1,0}^{r,s}$ in \cite{Simmons-Duffin:2016wlq}. }
\ba
G^{(d,a,b)}_{\t,\ell}(u,v)&=&\frac{\G(\t/2+\ell)^2}{\G(\t+2\ell)}\,
\sum_{k=0}^\infty\,u^{\t/2+k}\,
\sum_{n=0}^{2k}\, A_{k,n}\,f_{k,n}(v)\,,
\label{u-expansion-1-v}
\ea
where the basis functions 
\footnote{Note that $(1-v)^{\t/2+k}f_{k,n}(v)$ takes the same functional form as an $SL(2,\mathbb R)$ block 
parametrized by $\t/2+\ell+k-n$. 
It should be useful to make the substitution $u\rightarrow \tilde u \,(1-v)$, 
as one can systematically adapt the resummation techniques of the $SL(2,\mathbb R)$ blocks for the leading order terms to those at subleading orders. 
} 
are
\be
f_{k,n}(v)=(1-v)^{\ell-n}\,
{}_2F_1\bigg[\begin{matrix}
\t/2+\ell+k-n-a,\t/2+\ell+k-n+b\\
2(\t/2+\,\ell+k-n)
\end{matrix}\,;1-v\bigg]\,.
\label{f1}
\ee
At order $k$, the leading exponent of $(1-v)$ is $(\ell-2k)$, which decreases as $-2k$ due to the second order nature of the Casimir differential equation \eqref{Casimir-eq}. 
To derive the double lightcone expansion, we first perform a linear transformation. 
Then the small $u$ expansion reads:
\ba
G^{(d,a,b)}_{\t,\ell}(u,v)&=&v^{\frac{a-b}2}
\bigg[\frac{\G(b-a)}{(\t/2+\ell)_{-a}\,(\t/2+\ell)_{b}}\,
\sum_{k=0}^\infty\,u^{\t/2+k}\,
\sum_{n=0}^{2k}\,\tilde A_{k,n}\,\tilde f_{k,n}(v)
+(a\leftrightarrow b)
\bigg]\,,\nn
\label{u-expansion-v0}
\ea
where 
\be
\tilde A_{k,n}=\frac{(\t+2\,\ell)_{2(k-n)}}{(\t/2+\ell-a)_{k-n}(\t/2+\ell+b)_{k-n}}\,A_{k,n}\,,
\label{A-tilde}
\ee
\ba
\tilde f_{k,n}(v)=(1-v)^{\ell-n}\,
{}_2F_1\bigg[\begin{matrix}
\t/2+\ell+k-n+a,\t/2+\ell+k-n-b\\
1+a-b
\end{matrix}\,;v\bigg]\,.\quad
\label{f0}
\ea
The series coefficients of the double lightcone expansion \eqref{double-lightcone-expansion} can be expressed as sums of $\tilde A_{k,n}$:
\ba
C_{k_1,k_2}&=&\frac{\G(b-a)\,}{(\t/2+\ell)_{-a}\,(\t/2+\ell)_{b}}\sum_{n,m=0}^{k_2}\tilde A_{k_1,n}\,\frac{(n-\ell)_{k_2-m}}{(k_2-m)!}
\nn&&\qquad\qquad\times\,
\frac{( \t/2+\ell+k_1-n+a)_{m}\,( \t/2+\ell+k_1-n-b)_{m}}{m!\, (1+a-b)_{m}}\,.
\label{BigC-2}
\ea
Then we can compute the coefficients $A_{k,n}$ by matching $C_{k_1,k_2}$ in \eqref{BigC-2} with that in \eqref{BigC-1} order by order. 
At order $k$, we need to solve a system of $(2k+1)$ linear equations. 
The explicit solutions at low orders are
\be
A_{0,0}=1\,,\quad
A_{1,0}=\frac{(\t+\ell-1)_2\,\prod_{\a=\pm a,\pm b}(\t/2+\ell+\a)}
{(\t+2\ell-1)_2\,(\t+2\ell)_2\,(\t+\ell-d/2+1)}\,,
\nonumber
\ee
\be
A_{1,1}=\,-\frac{4ab\,\ell\,(\t+\ell-1)}{(\t-d+2)(\t+2\ell)(\t+2\ell-2)}\,,
\quad
A_{1,2}=\frac{\ell\,(\ell-1)}{2-d/2-\ell}\,,
\label{A-low-order}
\ee
which take simple factorized forms. 
From the concrete examples, we notice several interesting properties of $A_{k,n}$: 
\begin{itemize} 
\item
They are symmetric in $a^2$ and $b^2$. 
\item
They are proportional to $ab$ when $n$ is an odd integer, 
so the odd-$n$ cases vanish 
when two external operators have the same scaling dimension, i.e. $\D_1=\D_2$ or $\D_3=\D_4$.  
\item
If the spin $\ell$ is a non-negative integer, they always vanish when $n>\ell$. 
Then the $n$-summation terminates at $n=\ell$ at arbitrarily high orders
\footnote{At low orders, the $n$-summations terminate at $n=2k$ if $2k<\ell$.}, 
which is similar to the small $v$ expansion \eqref{small-v-solution-explicit}.  
\end{itemize}
To find a general expression, we decompose the low order coefficients into factorized rational functions as in the case of $B_{k,n}$. 
The dependence on $(a,b)$ is again particularly useful, 
which suggests the factorized building blocks should be symmetric in $\pm a, \pm b$. 
Then we are able to write these low order coefficients as double summations. 
In the end, we obtain a general formula for $A_{k,n}$: 
\ba
A_{k,n}&=&
\sum_{m_1,m_2=0}^{n_1}
(-1)^{n+m_1+1} \,4^{m_1+m_2}\,
\frac{(-\ell)_n\,(-n_1)_{m_1+m_2}(k-n_1+1/2)_{m_1}}
{n!\,m_1!\,m_2!\,(k-n+m_1)!}
\nn
&&\quad\times\,
\,\frac{(\t+\ell-1)_{2k-n}\,(-\t+d/2-\ell)_{n-k-m_1-m_2}\,(-\t+d-1)_{2(n_1-m_2)-n}}
{(\t+2\,\ell-n-1)_{2k-n}\,(\t+2\,\ell)_{2(k+m_1-n_1)-n}}
\nn
&&\quad\times\,
\frac{(1-d/2-\ell-k+n-m_1+m_2)
\,(3/2-d/2-\ell+n_2)_{m_2}
}
{(2-d/2-\ell)_{-k+n+m_2}\,(-1+d/2+\ell-m_2)_{k-n+m_1+m_2+1}}
\nn
&&\quad\times\,
\prod_{\a=\pm a, \pm b}\Big(\frac \t 2+\ell+\a\Big)_{k-n+m_1}\,\Big(\frac \g 2+\a\Big)_{m_2}
\times
\bigg\{\begin{matrix}
1\,\, & \text{if}\,\,\,\, n\equiv 0\\
4ab\,\, & \text{if}\,\,\,\,n\equiv 1
\end{matrix}\pmod{2}\, ,
\qquad
\label{Big-A}
\ea
where
we have introduced 
\be
n_1=\lfloor n /2\rfloor\,,\quad
n_2=\lfloor (n+1)/ 2\rfloor\,
\ee
to encode the minor differences between even and odd $n$ expressions,  
and $\lfloor x\rfloor$ is the floor function. 
Due to $(-n_1)_{m_1+m_2}$ and $1/(k-n+m_1)!$, there are additional constraints for nonzero summands:
\be
n_1\geq m_1+m_2\,,\quad
m_1\geq n-k\,,
\ee
which imply the general expression of $A_{k,n}$ becomes simpler when $n$ is close to $0$ or $2k$. 
\footnote{The simplicity of $A_{k,2k}$ and $A_{k,2k-1}$ was noticed earlier in \cite{Li:2017agi}. } 
Note that $a,b$ only appear in the last line of \eqref{Big-A}, 
and $m_1,m_2$ are the arguments of the associated Pochhammer symbols. 
One may notice that $A_{k,n}$ have poles at $\t+2\ell=m$ with $m=1,2,\dots$, but they are due to the basis functions \eqref{f1}. 
In Appendix \ref{appendix-1}, we discuss an alternative expansion without these spurious poles. 
As the concrete decompositions of $A_{k,n}$ are simpler than those of $B_{k,n}$, 
we need to study higher order terms to see the general pattern of the $(a, b)$ independent part. 
For large $k$, 
it is much more efficient to compute $A_{k,n}$ by matching the double lightcone expansion coefficients \eqref{BigC-2}
with \eqref{double-lightcone-expansion} 
than solving the Casimir differential equation \eqref{Casimir-eq}. 
Since our results here are based on the formulae in Sec. \ref{subsec-small-v}, 
we discuss the expansion in small $v$ before that in small $u$. 
We have also tested the formula \eqref{Big-A} to much higher orders using the Casimir equation \eqref{Casimir-eq} 
with rational parameterizations. 
After the substitution \eqref{zzb-definition}, the small $z$ or $\bar z$ expansion of \eqref{u-expansion-1-v} is also consistent with the closed-form expressions in $d=2,4,6$ dimensions \cite{Dolan:2000ut,Dolan:2003hv}.
\footnote{One should first set $\ell$ to a non-negative integer, and then set $d$ to an even integer. The two limits do not commute due to the singularities of $A_{k,n}$ at $d/2+\ell= \text{integer}$. 
If we take the even-$d$ limit first, then the closed-form expressions in \cite{Dolan:2000ut,Dolan:2003hv} 
are sums of \eqref{u-expansion-1-v} and its spin-shadow version. 
Furthermore, there are some typos in the last line of eq. (2.20) in \cite{Dolan:2003hv} for the $6d$ expression, 
where in the denominator $(\D+\ell-4)(\D+\ell-6)$ should be $(\D-\ell-4)(\D-\ell-6)$. }

Using the complete expression of the small $u$ expansion, we can also expand the conformal blocks 
around the fully crossing symmetric point \cite{Li:2017ukc} 
\be
u=v=1\,,
\ee
where the cross-ratios are invariant under all the crossing transformations, 
such as $1\leftrightarrow 2$ and $1\leftrightarrow 3$. 
\footnote{In terms of $(z,\,\bar z)$, the fully crossing symmetric point is at $z=e^{+ i \pi/3},\, \bar z=e^{- i \pi/3}$. }
As we are in the Euclidean regime, let us assume $\ell$ is a non-negative integer, so the $v\rightarrow 1$ limit is regular. 
The leading terms then read
\ba
G^{(d,a,b)}_{\t,\ell}(u,v)&=&\frac{\G(\t/2+\ell)^2}{\G(\t+2\ell)}\,
\sum_{k=0}^\infty\,\bigg[A_{k,\ell}+A_{k,\ell}\,(\t/2+k)\,(u-1)
\nn&&\qquad\qquad-
\Big(A_{k,\ell-1}+A_{k,\ell}\frac{(\t/2+k-a)(\t/2+k+b)}
{2(\t/2+k)}\Big)(v-1)+\dotsb
\bigg]\,.\qquad
\ea 
The fully crossing symmetric point is interesting 
in that all the crossing constraints can be systematically solved by expanding the correlator, i.e. $\mathcal G(u,v)$, around this point order by order.  
Then we only need to decompose the manifestly crossing symmetric correlator into physical conformal blocks \cite{Li:2017agi}. 

\section{Conclusion}
In summary, we have presented new analytic expressions of 4-point scalar conformal blocks in the lightcone expansions in \eqref{small-v-solution-explicit} and \eqref{u-expansion-1-v}, 
with closed-form coefficients in \eqref{big-B} and \eqref{Big-A}. 
They are explicit solutions of the Casimir differential equations in general dimensions for intermediate operators of arbitrary spin. 
They can be directly applied to the lightcone bootstrap to study the low twist spectra in general dimensions, 
and are particularly useful in $d\neq 2,4,6$ dimensions. 
Our results extend the analytic formulae for the lightcone limits to all the subleading terms, 
and should be useful for the systematic study of crossing equations at subleading orders of the lightcone expansion, 
which encode additional constraints and the information of higher twist operators. 

Using the lightcone expansions of conformal blocks, \eqref{small-v-solution-explicit} and \eqref{u-expansion-1-v}, 
one can compute the Lorentzian inversion in the lightcone expansion 
by changing the integration variables to $(u, v)$ or making the substitutions \eqref{zzb-definition}. 
\footnote{To carry out the inversion integral, 
it may be useful to perform a linear transform of the ${}_2F_1$ functions 
such that the argument $x$ becomes $x/(x-1)$. 
The Lorentzian inversion of a generic t-channel block with $a=b=0$ in the lightcone limit can be found in \cite{Li:2019dix}.}
As an infinite sum of the t-channel blocks over the spectrum can lead to divergences related to enhanced singularities, 
one may need to regularize the results properly. 
When the forms of enhanced singularities are known
\footnote{They are determined by the asymptotic behaviour of the CFT data at large spin.}, 
we can add and subtract the corresponding infinite sums to obtain convergent results \cite{Simmons-Duffin:2016wlq,Caron-Huot:2017vep}. 
More details on the regularization of the divergent sums are presented in Appendix \ref{appendix-2}. 
A different approach is to sum over the t-channel blocks before performing the lightcone expansion. 
For example, in the recent work \cite{Alday:2019clp}, 
a nontrivial double infinite summation of the t-channel blocks in general dimensions was partly carried out to extract some exact expressions in the lightcone limit. 
In \cite{Alday:2019clp}, the t-channel conformal blocks were computed order by order in $v$ using the Casimir equation. 
Our formula \eqref{u-expansion-1-v} provides general closed-form expressions to arbitrarily high order
\footnote{The small $u$ expansion of s-channel blocks is associated with the small $v$ expansion of t-channel blocks.}, 
so should be helpful for similar summations in the analytic conformal bootstrap.

Since spinning conformal blocks can be generated from the scalar blocks using differential operators, 
it would be interesting to revisit the spinning crossing constraints in the lightcone expansion, 
especially those associated with conserved currents and stress tensors \cite{Li:2015itl,Hofman:2016awc}.  

\begin{acknowledgments}
I am grateful to Johan Henriksson, Shinobu Hikami, Matthijs Hogervorst, David Meltzer, David Poland, Junchen Rong, Slava Rychkov and Hidehiko Shimada for 
enlightening discussions or conversations. 
I also thank Sriram Shastry for special encouragements. 
This work was supported by Okinawa Institute of Science and Technology Graduate University (OIST) 
and JSPS Grant-in-Aid for Early-Career Scientists (KAKENHI No. 19K14621).
This work was initiated during the Bootstrap 2019 conference at Perimeter Institute, 
which was supported in part by the Simons Collaboration on the Non-Perturbative Bootstrap.  
Research at Perimeter Institute is supported in part by the Government of Canada through the Department of Innovation, Science and Economic Development Canada and by the Province of Ontario through the Ministry of Economic Development, Job Creation and Trade. 
\end{acknowledgments}

\appendix
\renewcommand{\theequation}{\thesection.\arabic{equation}}
\addcontentsline{toc}{section}{Appendix}
\section*{Appendix}
\section{Alternative s-channel lightcone expansion}
\label{appendix-1}
Although the coefficients $A_{k,n}$ are quite simple at low orders, 
they have poles at $\t+2\ell=m$ with $m=1,2,\dots$, 
which are spurious and 
can be traced back to the basis functions $f_{k,n}$ with $n>k$. 
One can consider another set of basis functions to avoid these spurious poles. 
For example, we can expand the conformal blocks as
\ba
G^{(d,a,b)}_{\t,\ell}(u,v)&=&\frac{\G(\t/2+\ell)^2}{\G(\t+2\ell)}\,
\sum_{k=0}^\infty\,u^{\t/2+k}\,
\sum_{n=0}^{2k}\,\bar A_{k,n}\,(1-v)^{\ell-n}\,
\nn&&\qquad\qquad\times\,
{}_2F_1\bigg[\begin{matrix}
\t/2+\ell+k-n-a,\t/2+\ell+k-n+b\\
\t+2\,\ell+2k-n
\end{matrix}\,;1-v\bigg]\,,\qquad
\label{alternative-u-expansion}
\ea
where the parameter $\t+2\,\ell+2k-n$ in the basis functions is always greater than or equal to $\t+2\ell$. 
We can derive $\bar A_{k,n}$ from \eqref{double-lightcone-expansion}. 
It is easier to solve $\bar A_{k,n}$ than $A_{k,n}$ using the small $v$ expansion, 
as the exponents of the leading terms grow with $n$. 
We can solve the linear equations one by one, instead of $(2k+1)$ equations at the same time. 
The low order coefficients are
\be
\bar A_{0,0}=1\,,\quad
\bar A_{1,2}=\frac{\ell(\ell-1)}{2-d/2-\ell}\,,\quad
\ee
\be
\bar A_{1,1}=\ell
\bigg[\frac{2(a-b+d/2-1)-(\t+\ell-1)}{\t+2\ell}+
\frac{2(3-d-\ell)(\g/2-a)(\g/2+b)}{\g(2-d/2-\ell)(\t+2\ell)}
\bigg]\,,
\ee
\ba
\bar A_{1,0}&=&\frac{(\t/2+\ell-a)(\t/2+\ell+b)}{(\t+2\ell)(\t+\ell+1)}
\bigg[a-b+d/2-1
+\frac{(\g/2-a)(\g/2+b)}{\t/2+\g/2+\ell}
\nn&&\qquad\qquad\qquad\qquad\qquad\qquad\qquad\qquad
+\,\frac{\ell (3-d-\ell) (\g/2-a)(\g/2+b)}{\g(2-d/2-\ell)(\t/2+\g/2+\ell)}
\bigg]\,.\quad
\ea
In general, $\bar A_{k,n}$ is proportional to $(-\ell)_n$, so the $n$-summation also terminates for physical spin $\ell$. 
One can notice that the pole decomposition is similar to that of $B_{k,n}$ in \eqref{big-B}. 
After a linear transformation, a basis function in \eqref{alternative-u-expansion} is
\be
v^{(a-b)+n}\,(1-v)^{\ell-n}{}_2F_1
\bigg[
\begin{matrix}
\t/2+\ell+k+a,\t/2+\ell+k-b\\
1+a-b+n
\end{matrix};\,
v
\bigg]\,,
\ee
which is dual to \eqref{gkn-2}, 
together with a part where $a,b$ are properly interchanged. 
The $k$ dependence for small $n$ or $(2k-n)$ is not hard to guess, but we do not find a relatively simple expression for $\bar A_{k,n}$. 
In the $B_{k,n}$ case, the poles from $1/(1+a-b)_{m}$ are particularly helpful, but they are absent in the small $u$ expansion. 
Nevertheless, $\bar A_{k,n}$ can be expressed as sums of $A_{k,n}$ or $B_{k,n}$.

\section{Sums of t-channel conformal blocks}
\label{appendix-2}
In this appendix, we will discuss the potential divergences of the sums of t-channel conformal blocks near the lightcone. 
To simplify the discussion, we will assume the external scaling dimensions are identical, i.e. 
$a=b=0$.  
The t-channel conformal blocks become
\be
G^{(d,0,0)}_{\t,\ell}(v,u)
=
-\Big(\log u+2H_{\frac \t 2+\ell-1}+\pa_a\Big)
\sum_{k=0}^\infty\,u^{k}\,
\sum_{n=0}^\ell\,B_{k,n}\, g_{k,n}(v)\Big|_{a\rightarrow 0}\,,
\ee
where we have set $b=-a$ and the t-channel blocks are defined by the s-channel blocks with $u,v$ interchanged.

Let us first consider the correlator of generalized free fields:
\be
\mathcal G(u,v)
=\frac {u^{\D_\varphi}}{v^{\D_\varphi}}\,\mathcal G(v,u)
=\frac {u^{\D_\varphi}}{v^{\D_\varphi}}
\bigg(1+\frac {v^{\D_\varphi}}{u^{\D_\varphi}}+v^{\D_\varphi}
\bigg)\,.
\ee
In the double lightcone limit $v\ll u\ll 1$, 
the dominant contribution comes from the t-channel identity, i.e. $u^{\D_\varphi}/v^{\D_\varphi}$. 
In general, we also need to take into account the contributions from other t-channel operators. 
For generalized free fields, 
they are the t-channel conformal blocks of $\mathcal O_{n,\ell}=\varphi\Box^n \pa^\ell\varphi$ 
with $\D_{n,\ell}=2\D_\varphi+\ell+2n$. 
However, their sum, i.e. $1+u^{\D_\varphi}$, does not contribute to double discontinuities, 
which are associated with enhanced singularities in $v$. 
The Lorentzian inversion of the identity part gives the exact OPE data  
if we further restrict the spin $\ell$ to even integers. 

Nevertheless, let us consider these corrections to illustrate the potential divergences in 
the summations of t-channel blocks near the lightcone. 
The dominant term is $\mathcal G(v,u)=1+\dots$ from the identity operator. 
The leading corrections in $v$ are associated with the leading twist operators $\mathcal O_{0,\ell}$ with 
$\t=2\Delta_\varphi$.  
Their conformal blocks take the form $v^{\D_\varphi}(\log u+2H_{\t/2+\ell-1}+\pa_a)(1+\dots)$, 
where $(\dots)$ are non-negative integer powers of $u,v$. 
Assuming $\D_\varphi$ is positive, 
one finds that the sum over spin $\ell$ is not convergent. 
In addition, the $u$ dependence of each block is $u^k$ or $u^k\log u$, 
but the final result contains $u^{-\D_\varphi}$, which is more singular than $\log u$ in the lightcone limit $u\rightarrow 0$. 
In fact, the divergent sum of t-channel blocks over spin is closely related to the enhanced singularity in $u$. 
\footnote{To be more general, enhanced singularities are defined as terms that can be arbitrarily singular by acting the Casimir operator on them, 
so they are also called Casimir-singular terms. 
A convergent sum of $u^k$ or $u^k \log u$ can also generate enhanced singularities in $u$, which becomes a divergent sum only after applying the Casimir. }
Their relation is the crossing dual of that between the s-channel sum over spin and the enhanced singularity in $v$, 
captured by the double discontinuity from the t-channel identity.  

Now we consider the 3d Ising model, which is different from the generalized free theory. 
The t-channel corrections have non-zero double discontinuities. 
Let us focus on the correlator of the lowest $Z_2$-odd operator $\langle\s\s\s\s\rangle$ 
and the contributions of the leading twist family,
i.e. $\mathcal O_\ell\sim\s\pa^\ell\s$. 
Based on the spectrum of the Wilson-Fisher fixed points, 
we identify the lowest $Z_2$-even scalar $\e$ as the spin-0 operator 
and the stress tensor $T$ as the spin-2 operator. 
The numerical values of the OPE data of $(\e, T)$ were determined rather precisely \cite{Simmons-Duffin:2016wlq}. 
Using their data as the t-channel input, one can approximate the OPE data of the higher spin operators 
in the leading twist family, 
and the computation can be carried out using the Lorentzian inversion formula. 

For the complete t-channel contributions from the leading twist family, 
one should also take into account the higher spin operators. 
As explained in the generalized free case, 
the sum over spin is not convergent, 
but we know that they are related to enhanced singularity in $u$. 
Let us first consider the double lightcone limit
\be
\sum_{\ell=4,6,8,\dots}\,P_{\s\pa^\ell\s}\,G_{\s\pa^\ell\s}(v,u)\Big|_{u,v\rightarrow 0}
=
\frac {v^{\D_\varphi}}{u^{\D_\varphi}}
\bigg(1-
\sum_{\mathcal O=\e,T}\,P_{\mathcal O}\Big(\log v+2H_{\t_\mathcal O/ 2+\ell_\mathcal O-1}\Big)\,
u^{\t_\mathcal O/2}\bigg)+\dots\,,
\ee
where $(\dots)$ indicates the terms that are not enhanced singularities in $u$. 
Note that they are in one-to-one correspondence to the t-channel input for the Lorentzian inversion 
mentioned above. 
\footnote{According to the numerical values of $(\D_\varphi, \D_\e, \D_T)$, the $\mathcal O=\e$ part 
and the higher order terms in $u$ are not power law divergences in $u$. 
It is straightforward to include more operators and higher order terms.}
The small $v$ expansion of the enhanced singularities 
can be computed order by order using \eqref{u-expansion-1-v}, 
but a more efficient approach is to use twist conformal blocks \cite{Alday:2016njk}, 
which are sums of conformal blocks with identical twist. 
They satisfy a fourth order differential equation \cite{Alday:2016njk}, 
and the small $v$ limit plays the role of a boundary condition. 
The complete $v$ series at leading order in $u$ reads
\be
\frac {v^{\t/2}} {u^p}\rightarrow
\bigg(\frac {1-v} {u}\bigg)^p\,\bigg(\frac {v}{1-v}\bigg)^{\t/2}\,
{}_2F_1
\bigg[
\begin{matrix}
\g/2, \g/2\\
\g
\end{matrix};\,
-\frac v{1-v}
\bigg]\,,
\label{TCB}
\ee
where $\g=\t-d+2$ and the natural variables are $u/(1-v), v/(1-v)$. 
One can also derive the subleading enhanced singularities by solving the differential equation 
order by order in $u$. 
As in the generalized free case, 
these enhanced singularities in $u$ do not contribute to the double discontinuities, 
because they are multiplied by $u^{\D_\varphi}/v^{\D_\varphi}$ 
and the $v$-dependence is given by $v^k,\,v^k\,\log v$. 
It is the remaining part that contributes to nonzero double discontinuities in $v$. 

To compute the remaining part, 
we can use an identity for $SL(2, \mathbb R)$ blocks \cite{Simmons-Duffin:2016wlq} 
\ba
&&\sum_{\ell=0,1,2,\dots}\frac {(2h-1)\,\G(h+p-1)}{\G(p)^2\,\G(h-p+1)}\,
\frac{\G(h)^2}{\G(2h)}\,
(1-u)^h\,{}_2F_1\big(h,h,2h;1-u\big)
\nn&=&
\left(\frac {1-u} u \right)^p
+\frac {\G(\t/2+p-1)}{\G(p)^2\,\G(\t/2-p)}
\sum_{k=0}^\infty\,
\pa_k
\left(
\frac {\G(\t/2+k)}{(p+k)\,(k!)^2\,\G(\t/2-k-1)}\,\left(\frac u {1-u}\right)^k
\right)\,,\nn
\label{spin-sum-identity}
\ea
where $h=\t/2+\ell$. 
We can take the small $u$ limit and use this identity to regularize 
the sum of t-channel blocks in the double lightcone limit. 
In the second line of \eqref{spin-sum-identity}, 
the first term corresponds to an enhanced singularity in $u$, 
while the second term gives the exact difference between 
the divergent sum over spin and the enhanced singularity. 
For the complete enhanced singularities, 
we can multiply \eqref{spin-sum-identity} 
by some $v$-dependent functions, such as that in \eqref{TCB}. 
By adding and subtracting a linear combination of these identities, 
we obtain a convergent sum over spin. 
Therefore, we can use \eqref{small-v-solution-explicit} to directly compute the sums of t-channel conformal blocks order by order in $u$, 
which is more efficient than 
resumming the full $u$-dependent functions and then expanding the results in $u$.

\end{document}